\documentclass[aps,prl,twocolumn,showpacs,superscriptaddress,groupedaddress]{revtex4}  % for review and submission
\usepackage{graphicx}  % needed for figures
\usepackage{dcolumn}   % needed for some tables
\usepackage{bm}        % for math
\usepackage{amssymb}   % for math
\usepackage{soul}         % for strikethrough in draft

\begin{document}

% The following information is for internal review, please remove them for submission
%\widetext
%\leftline{Version 12 as of \today}

% the following line is for submission, including submission to the arXiv!!
%\hspace{5.2in} \mbox{Fermilab-Pub-04/xxx-E}

\title{Fast dynamics of radiofrequency emission in plasmas with runaway electrons}
%>>>\input author_list.tex       % D0 authors (remove the first 3 lines
                             % of this file prior to submission, they
                             % contain a time stamp for the authorlist)
                             % (includes institutions and visitors)
\author{P.~Buratti}
\email{paolo.buratti@enea.it}
\affiliation{ENEA, Fusion and Nuclear Safety Department, C.R. Frascati, Via E. Fermi 45, 00044
Frascati (Roma) Italy}
\author{W.~Bin}
\affiliation{ISTP-CNR, via R. Cozzi 53, 20125 Milano, Italy}
\author{A.~Cardinali}
\affiliation{ENEA, Fusion and Nuclear Safety Department, C.R. Frascati, Via E. Fermi 45, 00044
Frascati (Roma) Italy}
\author{D.~Carnevale}
\affiliation{Dip. di Ing. Civile ed Informatica, Università di Roma Tor Vergata, Italy}
\author{C.~Castaldo}
\affiliation{ENEA, Fusion and Nuclear Safety Department, C.R. Frascati, Via E. Fermi 45, 00044
Frascati (Roma) Italy}
\author{O.~D'Arcangelo}
\affiliation{ENEA, Fusion and Nuclear Safety Department, C.R. Frascati, Via E. Fermi 45, 00044
Frascati (Roma) Italy}
\author{F.~Napoli}
\affiliation{ENEA, Fusion and Nuclear Safety Department, C.R. Frascati, Via E. Fermi 45, 00044
Frascati (Roma) Italy}
\author{G.L.~Ravera}
\affiliation{ENEA, Fusion and Nuclear Safety Department, C.R. Frascati, Via E. Fermi 45, 00044
Frascati (Roma) Italy}
\author{A.~Selce}
\affiliation{ENEA, Fusion and Nuclear Safety Department, C.R. Frascati, Via E. Fermi 45, 00044
Frascati (Roma) Italy}
\author{L.~Panaccione}
\affiliation{ENEA, Fusion and Nuclear Safety Department, C.R. Frascati, Via E. Fermi 45, 00044
Frascati (Roma) Italy}
\author{A.~Romano}
\affiliation{ENEA, Fusion and Nuclear Safety Department, C.R. Frascati, Via E. Fermi 45, 00044
Frascati (Roma) Italy}
\author{FTU~Team}
\thanks{See the author list of G. Pucella {\sl et al.}, Overview of the FTU results, Nucl. Fusion {\bf 59}, 112015 (2019), https://doi.org/10.1088/1741-4326/ab19ef}
\affiliation{ENEA, Fusion and Nuclear Safety Department, C.R. Frascati, Via E. Fermi 45, 00044
Frascati (Roma) Italy}

\date{\today}

\begin{abstract}
Radiofrequency emission in the 0.4 - 3 GHz range from FTU plasmas in presence of runaway electrons (RE) has been measured with unprecedentedly high time resolution. Rapid emission bursts associated with enhanced RE pitch angle scattering reveal kinetic instabilities affecting evolution of the RE population from the buildup phase. Such measurements also provide a sensitive monitor for instabilities during early RE formation. The leading edge of radio bursts is much shorter than interleaving periods of low emission; spectral broadening during growth indicates nonlinear wave coupling as an explanation for the observed intermittency. Radiofrequency emission disappears at the beginning of post-disruption RE plateaus, and subsequently reappears in the shape of very intense bursts accompanied by strong rises of suprathermal electron cyclotron emission. 
\end{abstract}

\maketitle

%\section{\label{sec:level1}First-level heading}
% sections are not used for PRL papers

\textit{Introduction.}---Interactions between energetic particles and plasma waves play a role in various contexts, spanning from scattering of  runaway electrons (RE) in tokamaks  \cite{Breizman}, through local heating of radiation belt electrons \cite{VanAllen}, to cosmic ray acceleration in supernova remnants \cite{Amato}.

RE-driven wave instabilities are relevant for tokamaks since they can reduce RE beams energy. 
Runaway acceleration in tokamaks can occur either at low plasma density  (discharge startup in prticular), or at major disruptions, when the accelerating field is produced by plasma current quench. 
RE avalanches during tokamak disruptions are considered a major risk for plasma-facing components 
\cite{Breizman}. 
Collective interactions with the background plasma can enhance RE pitch angle scattering, leading to larger synchrotron losses, which in turn reduce both the critical electric field for avalanche multiplication 
\cite{Liu}
and the maximum energy of RE 
\cite{AB1}. 

Evidences of pitch angle scattering by kinetic instabilities can be traced back to early tokamak experiments 
\cite{Vlasenkov, Alikaev1, Alikaev2}. 
Instabilities were attributed to the anomalous Doppler effect, which allows energy transfer from RE motion along the ambient magnetic field to waves and gyromotion
\cite{Parail1, Coppi}. 
Relaxation cycles were observed, with duration much shorter than the interleaving periods, which raised the issue of rapid growth in spite of slow crossing of the stability boundary
\cite{Parail2}.

Direct evidences of radiofrequency waves involved in RE instabilities were reported in 
\cite{Alikaev2, Oomens} 
and more recently in 
\cite{Spong, Heidbrink}. 
Spectra consisting of multiple, coherent lines reported in 
\cite{Spong, Heidbrink}
were ascribed to whistler waves instabilities, and the possibility of injecting waves for RE mitigation was considered. 
Stability analysis pointed out the importance of collisional damping, and the existence of a threshold temperature
\cite{AB2}.

Measurements of wave emission at unprecedented time resolution are reported in this letter. Both continuous and bursting emissions are shown. Novel observations of spectral broadening during the leading edge of fast bursts are reported, which provide a direction for explaining rapid growth after stationary periods.
Radio burst accompanied by enhanced pitch angle scattering are found in different conditions, including post-disruption RE plateaus.

\textit{Experimental scenario and diagnostics.}---FTU is a high-field tokamak of major radius $R_0 = 0.935 \text{ m}$, with circular cross-section. The wall minor radius is 0.33 m; RE experiments were performed at 0.29 m limiter radius. Deuterium discharges are analyzed here, with plasma current up to $0.5 \text{ MA}$, and 4 or 5.3 T toroidal magnetic field.

Discharges with different RE content have been realized by operating at low gas fueling. The RE content obtained in this way is faint at densities above  $3 \times 10^{19} \text{ }m^{-3}$. 
Substantial RE populations at higher densities have been obtained by Deuterium pellet injection on low-density targets with very high RE content.
Post-disruption conditions with residual current carried by REs have been explored as well.

Hard X-ray (HXR) emission from in-flight RE collisions with plasma ions is measured by a collimated fast electron bremsstrahlung (FEB) camera with 5 ms time resolution and 20-200 keV energy range. Total HXR emission, including the one from RE impacts on plasma facing
components, is monitored with 50~\textmu s time resolution by a NE213 scintillator located outside the machine structure. 
The scintillator is sensitive to both HXR and neutrons. Neutron emission is measured by BF\textsubscript{3} proportional chambers and by fission chambers \cite{Bertalot}.
Fast events are spotted out by electron cyclotron emission (ECE) and Cherenkov probe diagnostics \cite{Causa, Bagnato}, with 10~\textmu s  and 0.65~\textmu s sampling time respectively. 

Radio waves generated by coupling with plasma waves have been collected by a wideband (log-periodic) antenna placed outside the vacuum vessel, in front of the exit of a vertical port closed by a dielectric window. 
The cutoff frequency due to propagation in the port is $\sim$400~MHz.
Variations of frequency spectra should reflect actual variations of plasma waves excitation, while absolute spectra are affected by unknown coupling and propagation losses. 
The antenna signal has been acquired by a NI PXIe-5186 fast digitizer, with 5~GHz analog bandwidth and 8-bit resolution. Data shown in this letter are acquired at 3.125 GS/s, corresponding to 0.2~s record duration, or at 6.25 GS/s (0.1~s record duration). 
A few pulses have been acquired with an analog spectrum analizer in place of the fast digitizer.

\begin{figure}
\includegraphics[scale=0.48]{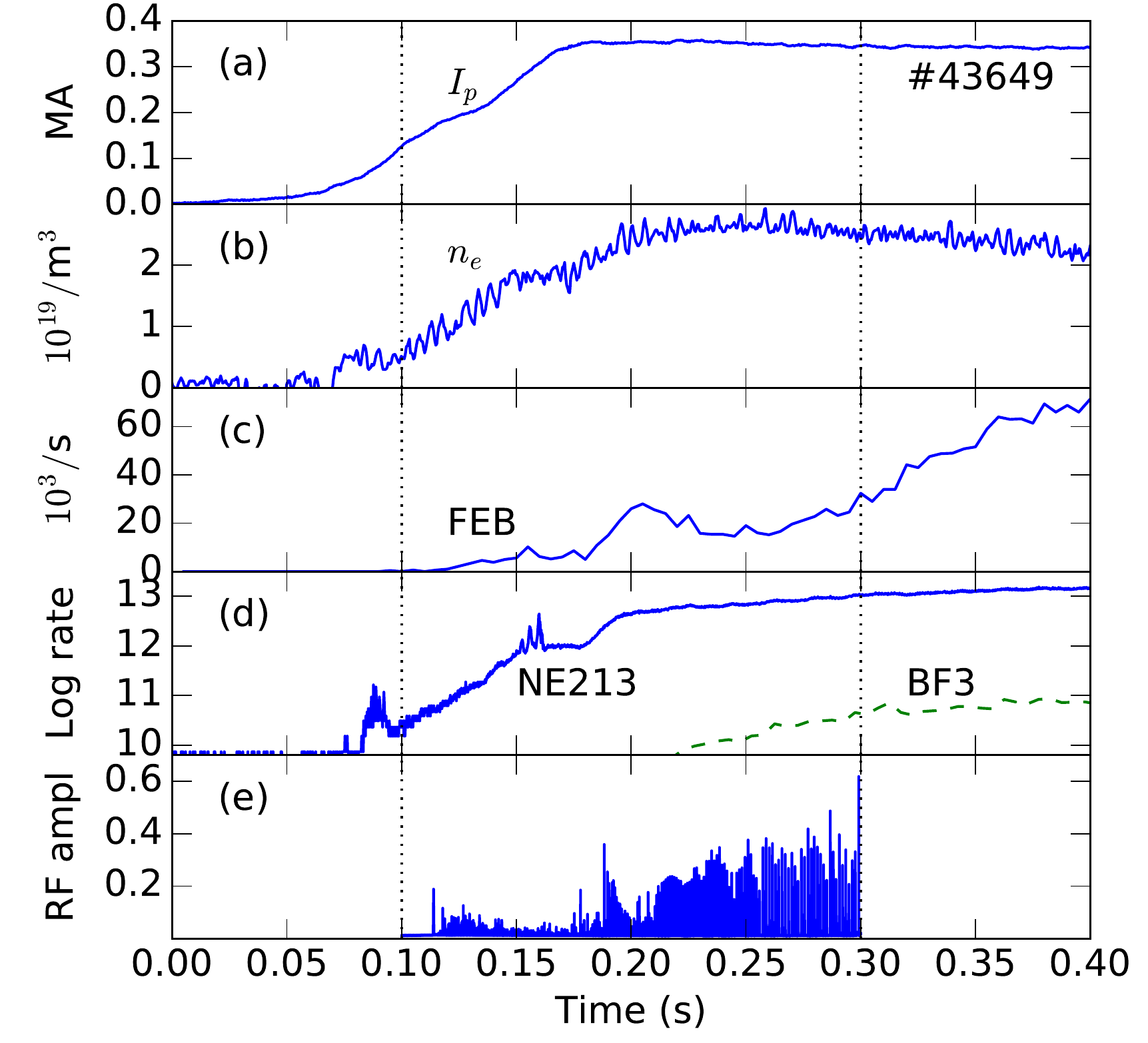}
\caption{\label{fig:fig_traces} Pulse 43649 at 5.3 T. (a) Plasma current. (b) Line-average density from a central chord of a CO\textsubscript{2} laser interferometer. (c) HXR count rate from an equatorial FEB channel. (d) Log\textsubscript{10} of HXR + neutron count rate from NE213 detector; the neutron contribution from cross-calibrated BF\textsubscript{3} detectors is shown by the dashed line. (e) Radiofrequency RMS amplitude normalized to digitizer saturation. Vertical dotted lines mark the acquisition interval.}
\end{figure}

\textit{Results}---Radiofrequency emission is routinely observable in pulses with runaway electrons. An example with HXR level in the middle of the explored range is shown in Fig.~\ref{fig:fig_traces}.
The fast digitizer acquisition interval includes part of current ramp-up and the beginning of current flat-top, Fig.~\ref{fig:fig_traces}(a). The peak electron temperature from Thomson scattering varies from 1.1 to 2.9 keV during this interval. 
Radio emission is intermittent; its amplitude, as estimated by moving RMS on 1~\textmu s intervals, is shown in Fig.~\ref{fig:fig_traces}(d). The first radio burst appears at 0.114~s, when HXR count rates as well as other RE markers are at very low levels. This indicates that kinetic instabilities influence RE dynamics already in the formation phase. 

\begin{figure}
\includegraphics[width=8.6cm]{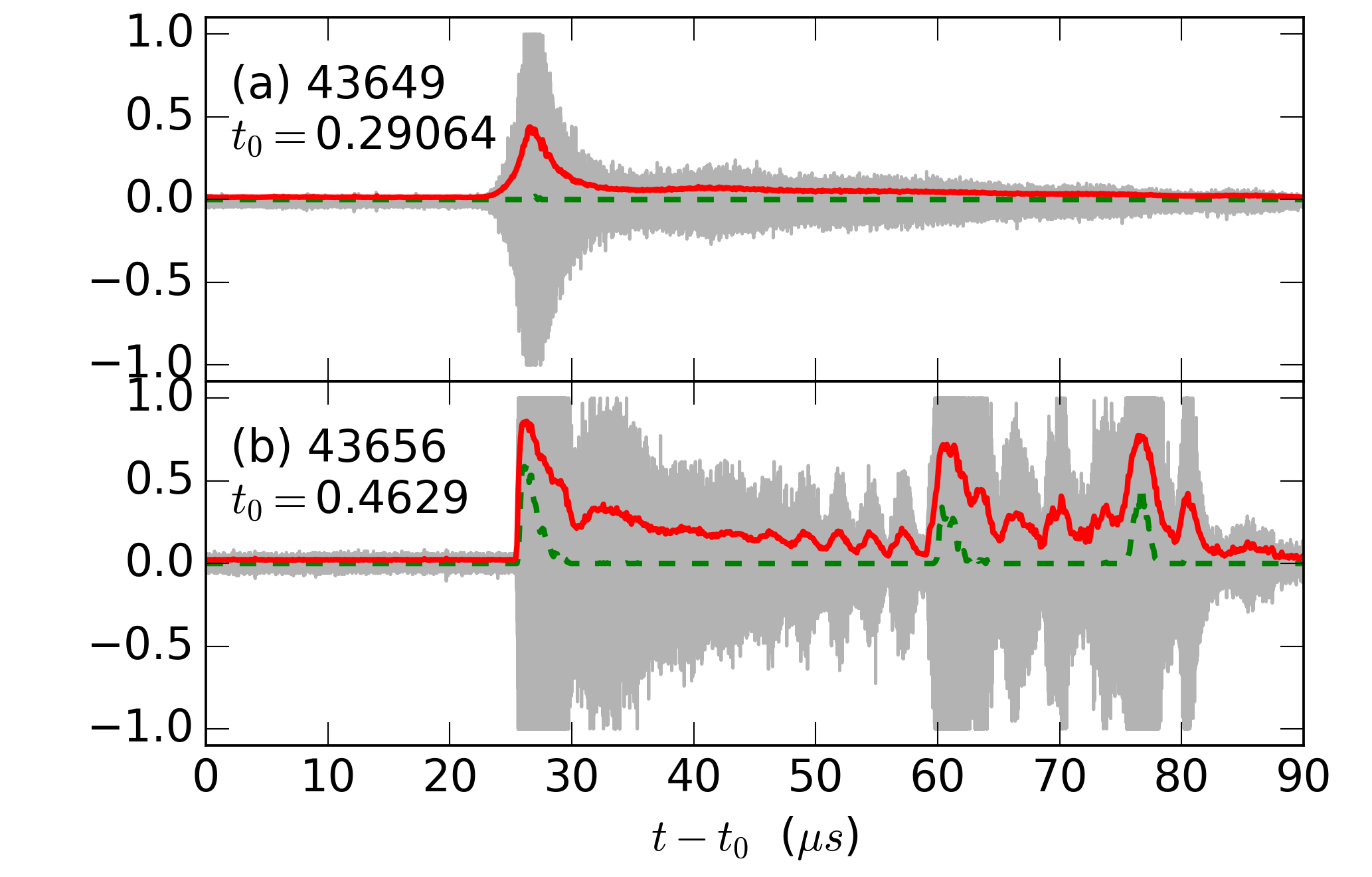}
\caption{\label{fig:fig_shapes_zoom} Raw signal normalised to saturation amplitude (grey), estimated amplitude by moving RMS on 100 ns intervals (red), and fraction of raw datapoints which saturate the digitizer dynamic range (green, dashed). 
(a) Pulse 43649, with $n_e=2.5\times10^{19}$. 
(b) Pulse 43656 with $n_e=1.5\times10^{19}$.}
\end{figure}

The shape of a single burst is shown in Fig.~\ref{fig:fig_shapes_zoom}(a). 
Amplitude grows exponentially for $\sim$5~\textmu s during the leading edge, with $1.1\times10^6 \text{ s}^{-1}$ growth rate. The period from the previous burst is 1.27~ms, which exceeds the e-folding time by three orders of magnitude. 
A radio burst from a pulse with higher RE content is shown in Fig.~\ref{fig:fig_shapes_zoom}(b). The leading edge is very sharp, and the tail structure is complex, in particular there are ringing oscillations, and a second sharp rising front appears after 35~\textmu s. 
Such a complex tail structure implies strong competition between driving and damping factors of the underlying instability.
Intermittency of radio emission is common to all examined pulses. No correlation between radio bursts timing and MHD activity is found, in particular there is no coincidence with sawtooth crashes when the latter are present.

\begin{figure}
\includegraphics[width=8.6cm]{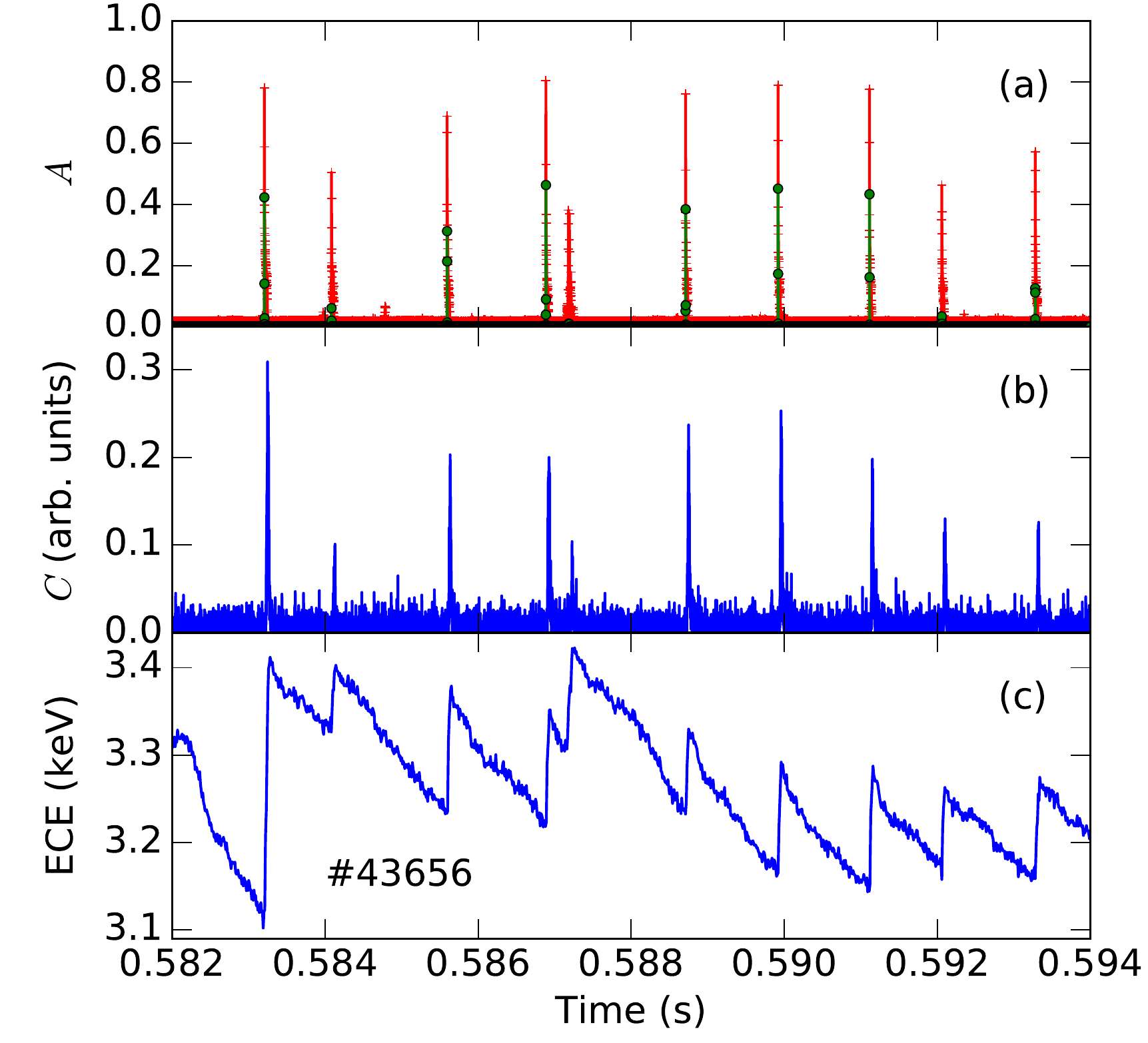}
\caption{\label{fig:fig_Cherenkov} Pulse 43656 at 5.3 T. (a) Normalized radiofrequency fluctuation amplitude on 1 \textmu s intervals (red crosses) and fraction of saturated raw datapoints (green dots); (b) Cherenkov probe signal; (c) ECE radiation temperature at 370 GHz. The average ECE level (above 3 keV) is strongly suprathermal; rapid variations at radio bursts range from 2\% to 10\%.}
\end{figure}

Rapid variations of the electron distribution function are monitored using an ECE channel and a Cherenkov probe. The ECE channel lies between the second and the third harmonic of the electron gyrofrequency, where radiation temperature is $\sim$0.1 keV if RE are absent.  Increasing ECE is a sign of increasing RE perpendicular momentum. The Cherenkov probe detects escaping RE at an equatorial location in the limiter shadow.
Most radio bursts coincide with rapid ECE signal rises, as well as with bursts of the Cherenkov signal, as exemplified in Fig.~\ref{fig:fig_Cherenkov}.
Radio emission can be envisaged as a useful diagnostic, in that it allows spotting out this kind of events even in cases with faint RE content and hardly detectable ECE and Cherenkov signal variations.

\begin{figure}
\includegraphics[scale=0.48]{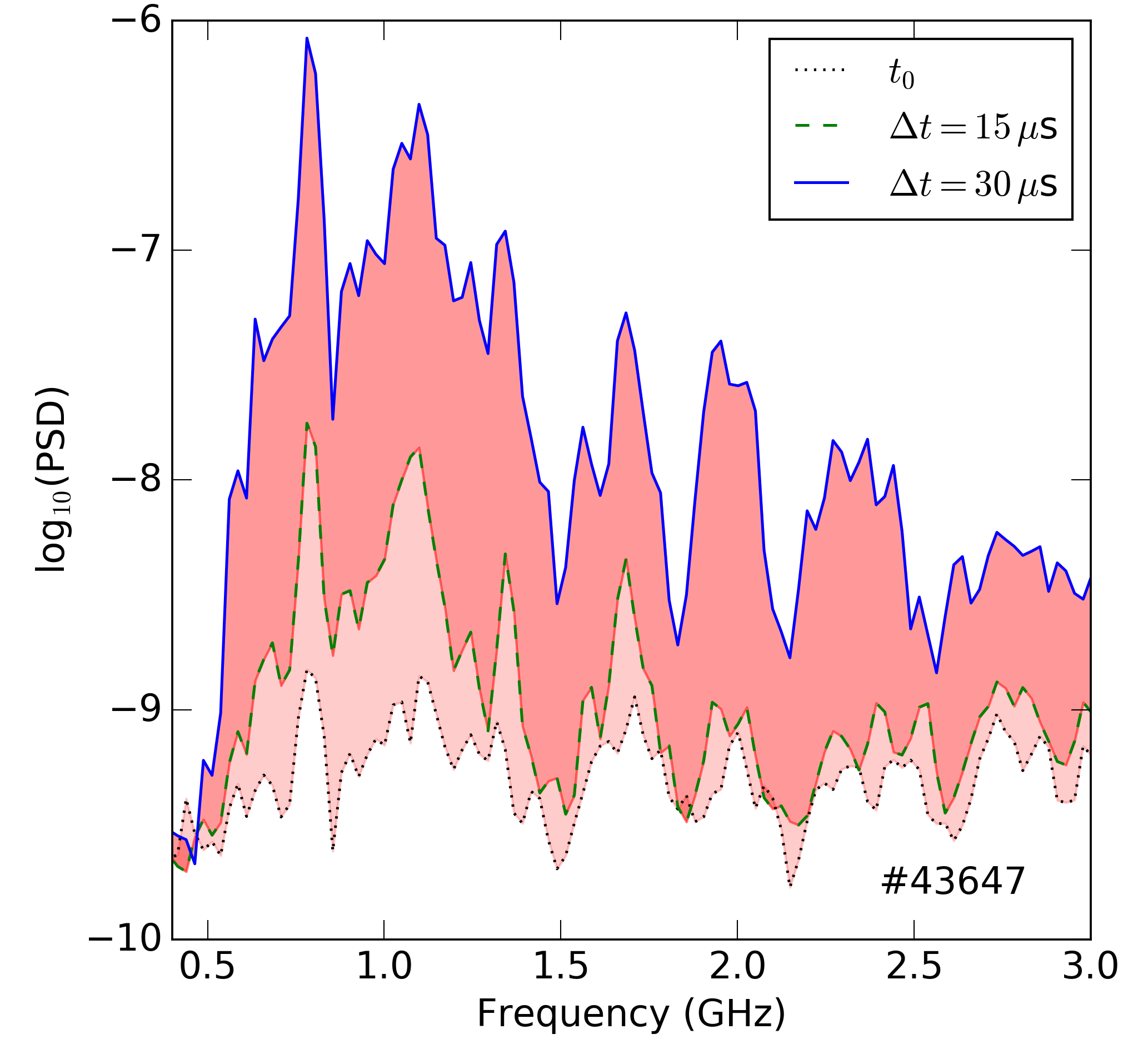}
\caption{\label{fig:fig_spectral} Welch's PSD (on 20 intervals of 256 points), just before a radio burst ($t_0=0.122435$ s, lower curve), during early exponential growth ($t-t_0=15$ \textmu s, curve in the middle) and at maximum growth rate ($t-t_0=30$ \textmu s, upper curve), for pulse 43647 at 4 T.}
\end{figure}

Evolution of frequency spectra during the leading edge of a radio burst is shown in Fig.~\ref{fig:fig_spectral}. 
Amplitude keeps growing for $\sim$30~\textmu s in this case, with $\sim$10$^5 \text{ s}^{-1}$ growth rate. 
The initial spectrum shown by a dotted line in Fig.~\ref{fig:fig_spectral} is representative of the nearly constant power spectral density (PSD) observed during low-emission phases between bursts.
The dashed line shows a PSD during the early growth phase, when amplitude is $\sim$13\% of the maximum one. Significant changes occur in a relatively narrow range, from 0.6 to 1.7~GHz, as emphasized by the lighter shading in Fig.~\ref{fig:fig_spectral}. The third spectrum is taken 15~\textmu s after the second one, when amplitude is at $\sim$60\%; the darker shading highlights substantial broadening with respect to the early growth phase. 
Spectral broadening occurring on a time scale comparable to the amplitude e-folding time is indicative of strongly nonlinear wave coupling, which is a possible mehanism for rapid growth after relatively long stationary periods.

\begin{figure}
\includegraphics[width=8.6cm]{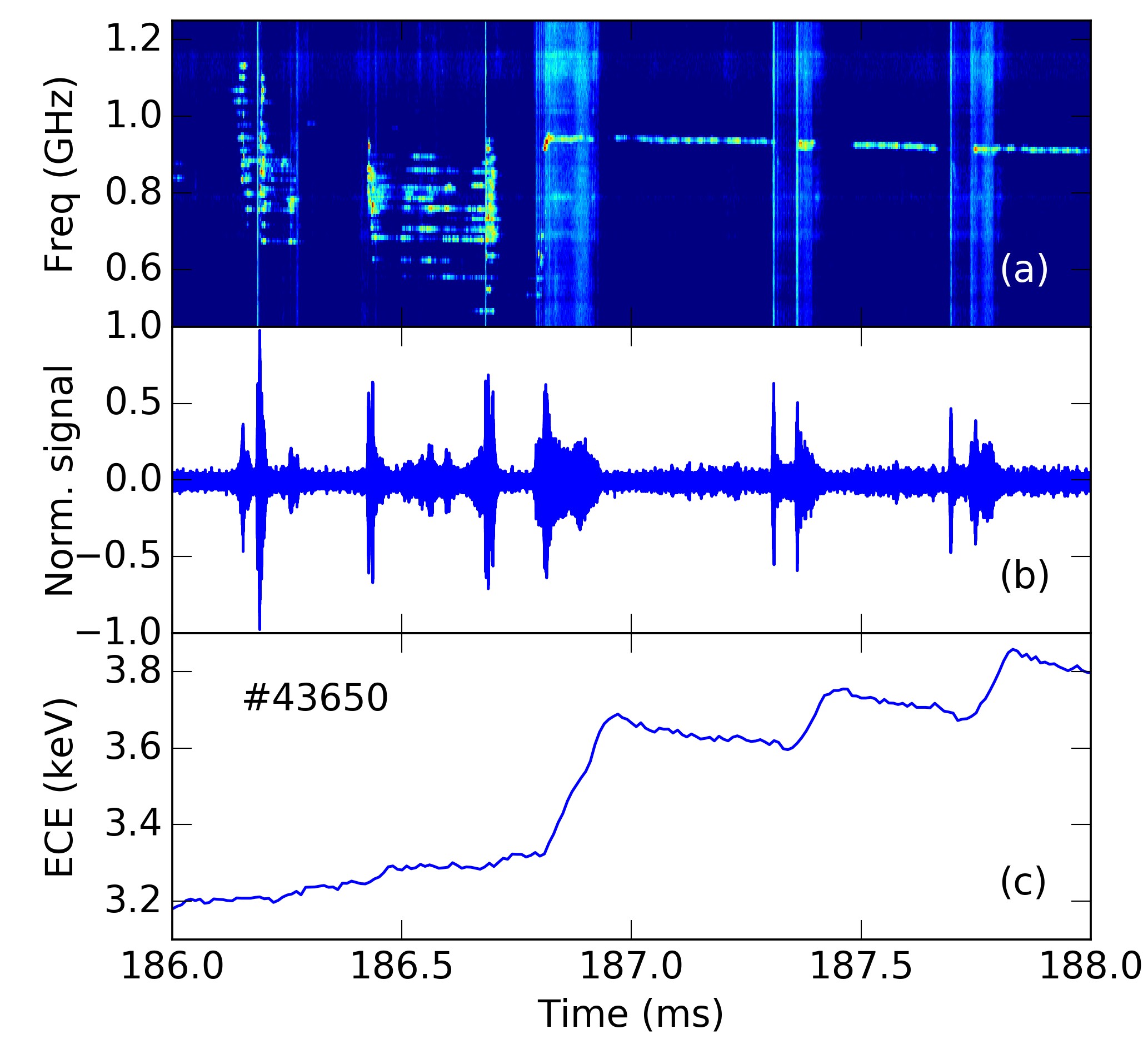}
\caption{\label{fig:fig_multilines1} (a) Spectrogram of radiofrequency emission from Welch's PSD on 16 semi-overlapped intervals with 2048 points each; (b) raw signal normalized to digitizer saturation amplitude; (c) ECE radiation temperature at  370 GHz.}
\end{figure}

Frequency spectra of radio bursts are mostly broad, while discrete lines are observed occasionally. An example with multiline excitation, followed by broadband emission periods alternating with single-line spectra, is shown in  Fig.~\ref{fig:fig_multilines1}. The ECE signal slope only increases during periods of broadband emission. 

\begin{figure}
\includegraphics[width=8.6cm]{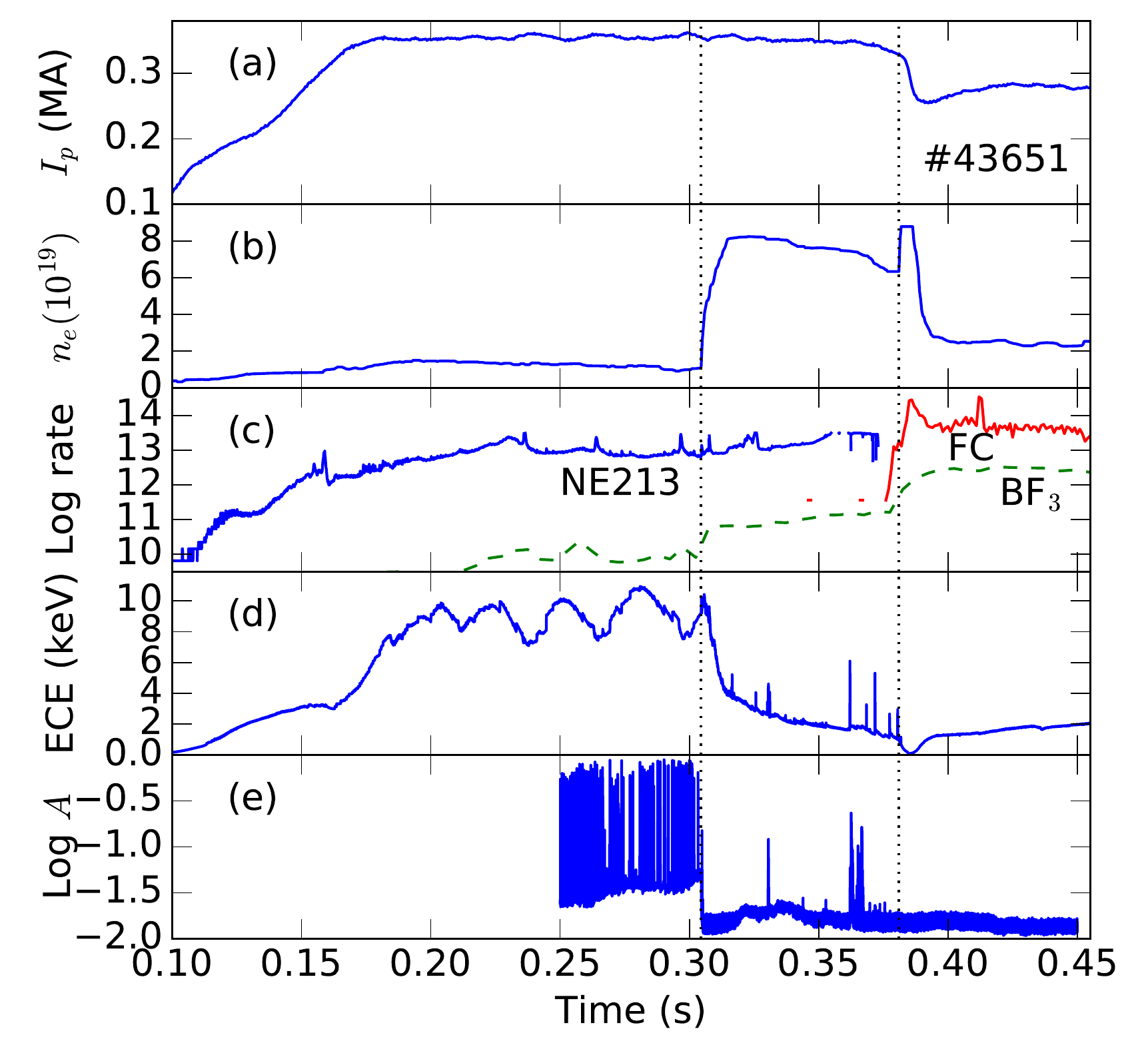}
\caption{\label{fig:fig_postpellet} Time evolution of a discharge with pellets injected at 0.304 s and 0.38 s (vertical dotted lines). (a) Plasma current. (b) Average density. (c) Log\textsubscript{10} of: count rates of NE213 scintillator (not shown when saturated), average of two fission chambers close to the poloidal limiter (FC) and average of BF\textsubscript{3} detectors. (d) ECE radiation temperature at 370 GHz; (e) Log\textsubscript{10} of normalized radiofrequency amplitude; the record ends at $t=0.45$ s.}
\end{figure}

Plasmas with substantial RE content and relatively high density can be obtained by injecting a large Deuterium pellet in low-density pulses with high RE content, as shown in Fig.~\ref{fig:fig_postpellet}.
Pre-pellet density is low and RE markers are at high levels. Intense radiofrequency bursts superposed to a continuous emission level are present before pellet injection, see Fig.~\ref{fig:fig_postpellet}(e).
Density increases from 1.2 to $8.3\times10^{19}$ (60\% of the Greenwald limit) after pellet injection at $t=0.304\text{ s}$, as shown in Fig.~\ref{fig:fig_postpellet}(b). 
Density keeps increasing for 10 ms after injection, because of incomplete pellet ablation.
Central temperature from Thomson scattering decreases from 2.4 to 0.5 keV.

A substantial RE population survives pellet injection, indeed total HXR emission remains high, Fig.~\ref{fig:fig_postpellet}(c).
Radio bursts disappear and the baseline emission level reduces by factor five within 0.9 ms after pellet injection.
ECE declines, likely because perpendicular momentum of RE is no longer fed by pitch-angle-scattering instabilities, but its level remains suprathermal, as shown in Fig.~\ref{fig:fig_postpellet}(d). 

\begin{figure}
\includegraphics[width=8.6cm]{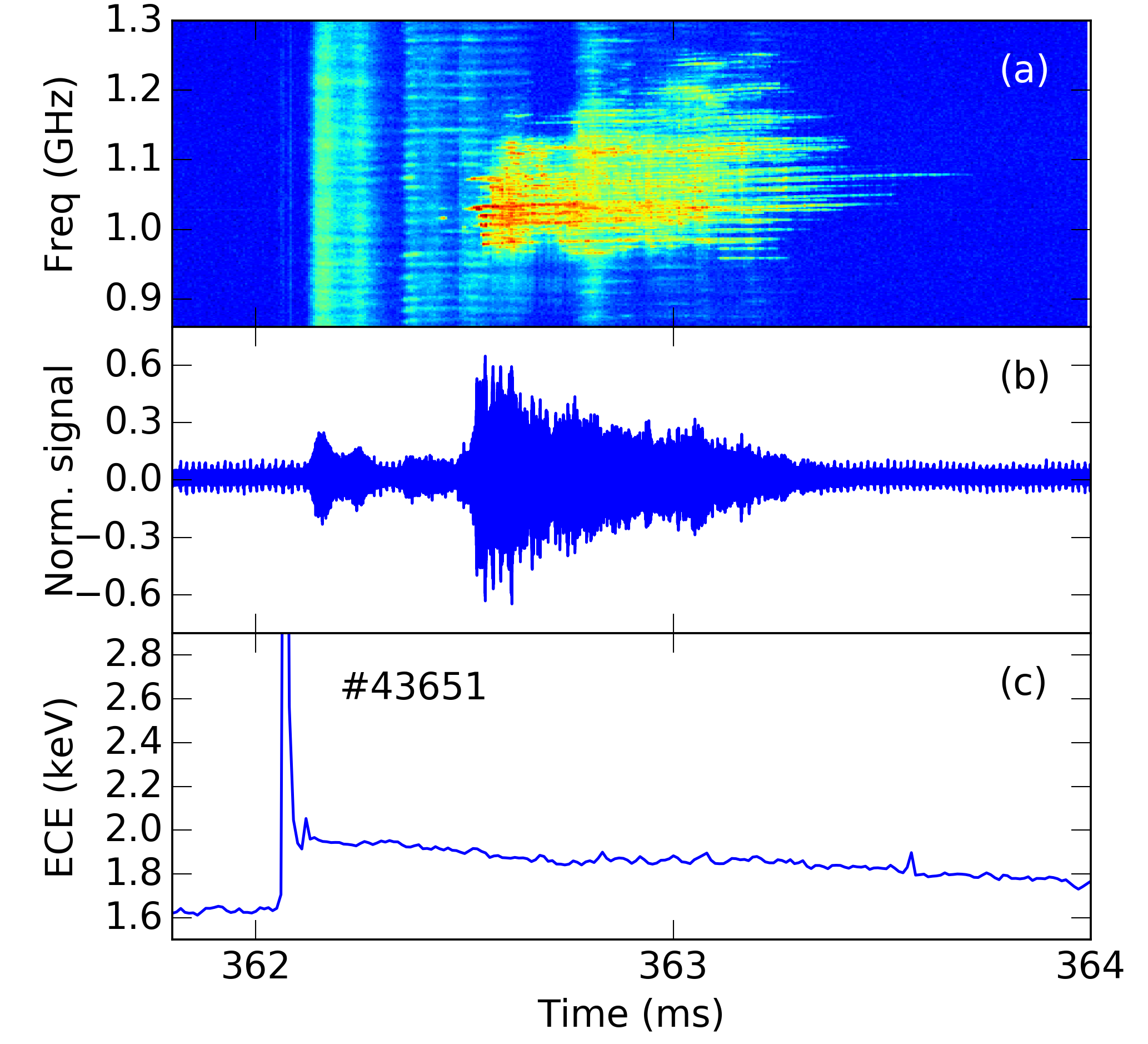}
\caption{\label{fig:fig_multilines2} (a) Spectrogram of radiofrequency emision from Welch's PSD on 16 semi-overlapped intervals with 2048 points each; (b) raw signal normalized to digitizer saturation amplitude; (c) ECE radiation temperature at  370 GHz. The ECE spike is out of scale.}
\end{figure}

After pellet injection, the RE content increases and sporadic instability evidences reappear. The aspect of such instabilities differs from the one observed at low density, in particular there are spikes instead of small steps in ECE; furthermore, some ECE spikes have no counterpart in radiofrequency emission. Spectral analysis of one such event is shown in Fig.~\ref{fig:fig_multilines2}. Emission is dominated by a relatively narrow 0.95-1.15 GHz band, with a fine substructure which resembles the ones reported in \cite{Spong, Heidbrink}. 

The sudden emission drop after pellet injection is likely due to collisional damping, as the electron-ion collision frequency increases by two orders of magnitude (to $\sim$$10^6$ s$^{-1}$).
The density variation also modifies plasma waves dispersion, and consequently the RE-excited wave branches. 
The ratio between electron cyclotron and plasma frequencies (as estimated with central magnetic field and line-average density) is  $\omega_{ce}/\omega_{pe}=4.8$ before pellet injection; it follows that, according to the cold plasma dispersion relation, only RE with relativistic factor $\gamma>50$ can be in anomalous Doppler resonance with whistler modes, while lower energy RE interact with magnetized plasma oscillations. The frequency ratio reduces to $\omega_{ce}/\omega_{pe}=1.8$ after pellet injection, and interaction with whistler modes can occur at lower energies, $\gamma>8$.

\begin{figure}
\includegraphics[width=8.6cm]{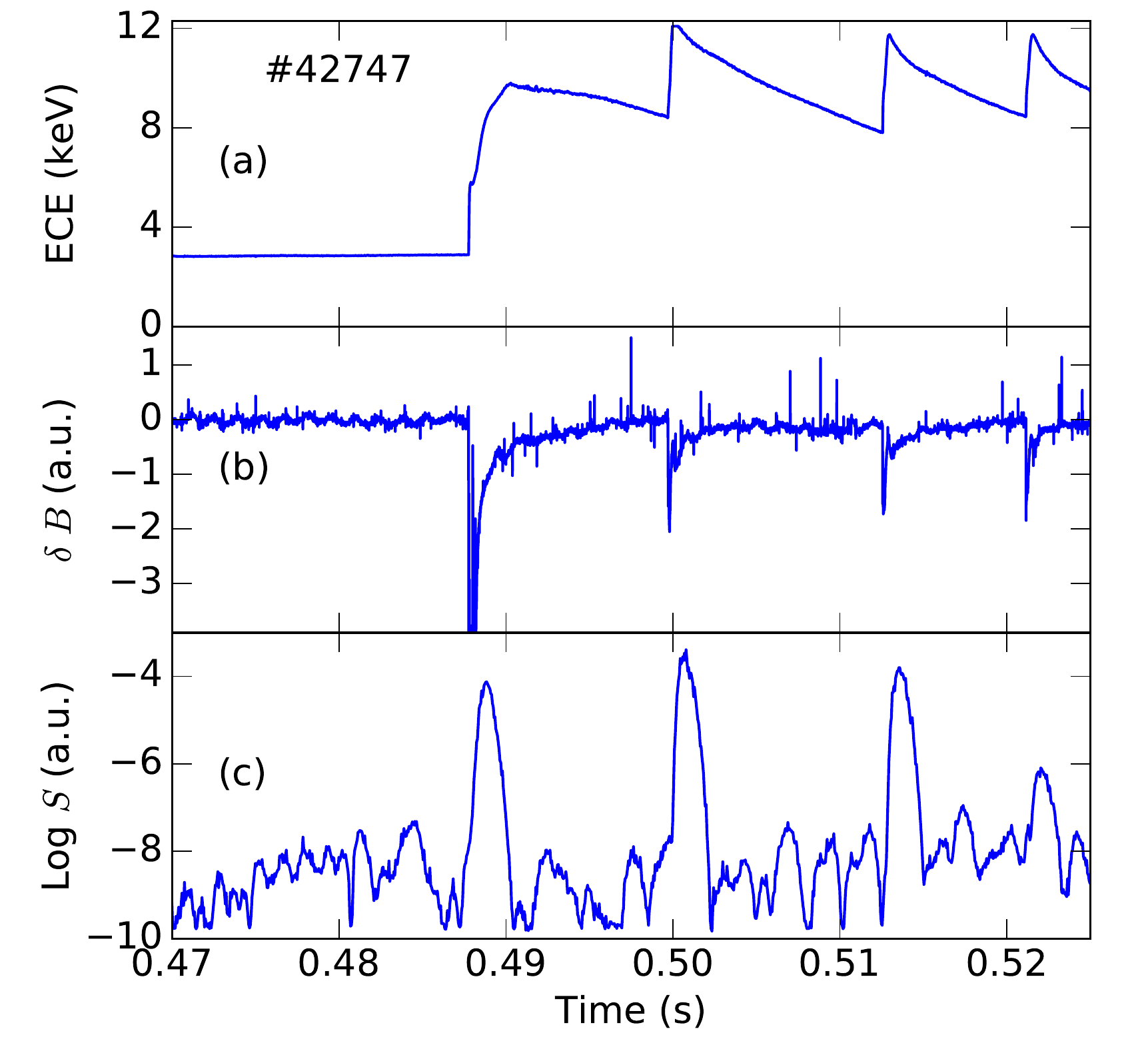}
\caption{\label{fig:fig_radiometer} Development of kinetic instabilities during a RE plateau after a quiescent phase (which lasts for 0.2 s in this case).  (a) ECE radiation temperature at 370 GHz. (b) Magnetic pickup signal; the first spike is out of scale. (c) Spectrum analizer readout, with frequency tuned at around 500 MHz.}
\end{figure}

A partial current quench followed by a plateau can be seen in Fig.~\ref{fig:fig_postpellet}(a) after the second pellet injection. The background plasma temperature drops below the Thomson scattering operating range as the second pellet is injected. RE energies increase above $\sim7$~MeV, as shown by a substantial neutron rate produced by photonuclear reactions, the fusion contribution being negligible at low temperature. Radio emission and ECE traces are smooth during this phase, while ECE spikes (not shown) reappear after the end of the fast digitizer record. Quiescent phases like this one are typical of post-disruption RE plateaus. 
Strong collisional wave damping due to plasma cooling can justify the extinction of kinetic instabilities.
However, such quiescent phases are often terminated by sequences of violent instabilities; this happens if the RE beam lasts long enough ($>0.2$ s typical) in pulses with position-controlled current ramp-down \cite{Carnevale}.  
Large ECE rises appear, which are accompanied by magnetic perturbations, as shown in Fig.~\ref{fig:fig_radiometer}. These observations are akin to the ones reported in \cite{Vlasenkov, Alikaev1, Alikaev2}. 
Instead, events shown in Figs.~\ref{fig:fig_shapes_zoom} and \ref{fig:fig_Cherenkov} have no detectable counterpart in magnetic signals; apparently, magnetic perturbations are absent if the background plasma is hot and carries most of the plasma current.
Radiofrequency activity is visible in the readout of the analog spectrum analizer, Fig.~\ref{fig:fig_radiometer}(c), while the fast digitizer is completely saturated.
Resurgence of kinetic instabilities during RE plataeus has been attributed to progressive reheating of the background plasma \cite{Carnevale}.

\textit{Conclusions}---
Bursts of radiofrequency emission have been systematically observed in conjunction with kinetic instabilities that lead to pitch angle scattering of runaway electrons. 
Impulsive pitch angle scattering events have been spotted out already in the early formation stage of the RE population, showing the importance of kinetic instabilities in all phases of RE life. 
It is worth noting that measurements on FTU have been performed at $\omega_{ce}/\omega_{pe}\approx5$, in the ballpark of ITER startup conditions. 

Spectral broadening on the time scale of the amplitude e-folding time has been observed, which indicates that the rapid onset of RE instabilities after relatively long quiescent periods is due to non-linear wave coupling. Further indications on strongly non-linear behavior are provided by ringing oscillations in the time domain.

Very strong radio bursts have been observed during RE current plateaus, showing that instability conditions can be met in spite of strong collisional damping that is typical of post-disruption plateau conditions.

%>>>\input acknowledgement.tex   % input acknowledgement

\end{document}